\documentclass[runningheads]{llncs}

\input{psfig.sty}

\begin{document}

\pagestyle{headings}

\mainmatter

\title{Computer Science in Physics}

\titlerunning{Computer Science in Physics}

\author{A. P. Young}

\authorrunning{A. P. Young}

\institute{Physics Department\\
University of California \\
Santa Cruz,  CA 95064 \\
\email{peter@bartok.ucsc.edu}\\
\texttt{http://bartok.ucsc.edu/peter}
}

\maketitle

\begin{abstract}
This talk describes how techniques developed by Computer Scientists have
helped our understanding of certain problems in statistical physics which
involve randomness and ``frustration''.  Examples will be given from two
problems that have been widely studied: the ``spin glass'' and the
``random field model''.

\end{abstract}

\section{Introduction}

An important part of the area of physics known as ``statistical physics'' is
the study of phase transitions, at which the system converts from one state to
another. Most interest has centered on ``second order'' or ``continuous''
transitions, in which the property which distinguishes the two phases vanishes
continuously as the transition is approached. The disappearance of the
magnetization of a ferromagnet, such as iron, as the temperature is increased is
generally continuous. At the other type of transition, known as ``first
order'' or ``discontinuous'', there is a jump in the
properties of the system as the transition is crossed,
and also a
latent heat. An everyday example of a first order transition is the freezing
of water. 

We shall focus on magnetic transitions in this talk because (i) they can be
represented by simple models amenable to numerical study, and (ii) there are
many experimental systems which are describable by these models.
One of the major advances in the field has been the realization that 
behavior in the vicinity of the transition (called the ``critical
point'') is ``universal''~\cite{univ}.  This means that ``critical
behavior'' does not depend on the microscopic details of the system
but only on much more
basic features such its dimensionality and symmetry. Consequently one can use
relatively simple models, which can be readily simulated, to make precise
comparisons with experiment. That said, it should be emphasized that
universality is much better justified for clean systems than for the systems
with disorder which we shall be considering in this talk. One goal of applying
sophisticated algorithms from computer science to these problems will be to
see if universality also holds for disordered systems.

The simplest model which describes a magnetic transition, known as the Ising
model, has a variable at each site on a regular lattice which can point
either ``up'' or ``down''.  This represents the orientation of the magnetic
moment of an atom, and is simplified to only allow two possible orientations.
We shall follow standard notation in calling these variables ``spins'',
labeled $S_i$ where $i$ denotes a lattice site. It is convenient to denote
the up spin state by $S_i = 1$ and the down spin state by $S_i = -1$. 

Neighboring sites on the lattice interact with each other. If the
interaction favors parallel alignment of the spins, it is called
``ferromagnetic'', while an interaction favoring anti-parallel alignment is
called ``anti-ferromagnetic''. The energy (confusingly called the
``Hamiltonian'' in the physics literature) can therefore be written as
\begin{equation}
E = -\sum_{\langle i, j \rangle} J_{ij} S_i S_j ,
\label{ham}
\end{equation}
where the sum is over all nearest neighbor pairs of the lattice (counted once)
and the interactions are labeled $J_{ij}$. If all the $J_{ij}$ are positive
then the state of lowest energy (ground state) has all spins parallel (either
all $+1$ or all $-1$) and is called a ferromagnet. As the temperature is
increased, the net alignment of the spins, the magnetization, decreases and
vanishes at a critical temperature $T_c$, where thermal noise, which tends to
randomize the spins, overcomes the interactions, which tend to make them
order.

The situation with all interactions negative is simple if the sites of the
lattice can be divided into two sublattices, A and B, such that all the
neighbors of A are in B and vice-versa. Such lattices are said to be
``bi-partite'', and are the only type that we shall consider here. A square
grid is an example of a bipartite lattice. The ground state for a bipartite
lattice with negative interactions has all spins $+1$ on sublattice A and all
spins $-1$ on sublattice B, or vice-versa. Such a state is called an
antiferromagnet. Again the ordering decreases to zero at a critical
temperature.

The problems of interest to us have two more ingredients. The first is
{\em disorder}\/.
The interactions are not all equal but are chosen in some random
way. The simplest model for disorder
is to pick each interaction from a probability distribution,
independently of all the others.  The second ingredient is
``{\em frustration}\/'' or
competition between the interactions. For the model in Eq.~(\ref{ham}) this
can be incorporated by allowing the {\em sign}\/ as well as the magnitude of
the interaction to be random.

\begin{figure}
\centerline{\psfig{figure=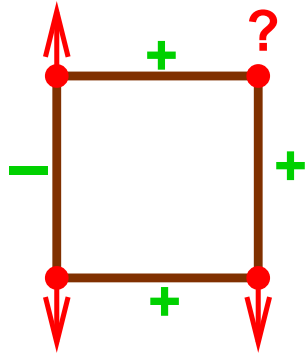,height=5cm}}
\caption{
A simple example illustration frustration as discussed in the text.
}
\label{fig:square}
\end{figure}

That this leads to frustration can be seen in the ``toy'' example in
Fig.~\ref{fig:square}.
This shows just four sites round a square with one anti-ferromagnetic
(negative) interaction and three ferromagnetic (positive) interactions. If the
spins along the bottom row and top left corner are oriented in directions which
minimize the energy (as shown) the spin at the top right is ``frustrated''
since it receives conflicting instructions from its two neighbors. It wants to
be parallel to both of them which is impossible.
It is easy to see that there is frustration if there is an
odd  number of negative interactions round the square, which is then called a 
``frustrated square''. 

If we extend this toy example of four sites to a large lattice,
choosing the sign (and possibly the magnitude) of the interactions at random,
determining the ground state is non-trivial, which is not the case if
all interactions have the same sign. This type of problem, which has been
extensively studied, is called a ``spin glass''~\cite{book,by}. There are
many experiments on magnetic systems with these features of disorder and
frustration but it would take us too far afield to discuss them here. Spin
glasses are also considered as prototypes for other systems with frustration
and disorder which have many features in common. Examples
are neural networks, protein folding models, elastic manifolds in random
media, and the ``vortex glass'' transition in superconductors in a magnetic
field. These are discussed in the articles in Ref.~\cite{book}

Determination of the ground state in systems with disorder and frustration
is an optimization problem, in which the
``cost function'' that has to be minimized is the energy. As we shall see,
algorithms from computer science enable us to calculate the ground state of
spin glasses for surprisingly large lattice sizes, at least in certain cases. 
An excellent introduction to optimization algorithms as applied to problems in
physics is the review by Rieger~\cite{heiko}.

Spin glasses and other problems with disorder and frustration are hard because
the energy varies in a complicated manner as one moves through configuration
space. There are local minima of the energy, which we will call ``valleys'' in
the ``energy landscape'', separated by ``barriers'' (i.e. saddle points).
Different local minima can have similar energies but have very different
configurations of the spins.  At a finite temperature the systems should spend
time in different valleys with relative proportions given by the appropriate
Boltzmann factors~\cite{boltz}. Only one of the local minima will be the global
minimum (ground state). This can be hard to find if there are many minima
and/or the global minimum has a small ``basin of attraction''. However, it is
generally quite easy to find a minimum with energy close to the ground state
energy, for example by the method of ``simulated annealing''~\cite{siman}.

The precise value of the ground state energy will depend on the particular
choice of the random interactions (remember they were picked from a
distribution). In physics we usually look at ``intensive'' quantities
(those which do not depend on size of the system, $N$, as $N \to \infty$) such
as the ground state energy {\em per spin}\/. Many intensive quantities
are ``self-averaging'' which means that its value does not depend
on the realization of disorder for $N \to \infty$. However, there {\em  are}
sample-to-sample fluctuations, generally of order $1/\sqrt{N}$, for
finite-sized systems, so we need to average results over many
realizations of disorder. This makes the problem even more computationally
challenging than if we just had to solve for one sample, but fortunately
averaging over samples is clearly
``trivially parallelizable'', so we can easily take advantage of the
large-scale parallel machines that are widely available at present, or just run
the code on a ``farm'' of independent workstations. 

In this talk I will also discuss another widely studied problem with
frustration and disorder, known as the
random field model. A magnetic field will prefer a spin to align in one
direction rather than the other, and so can be represented in the expression
for the energy  by terms linear in
the spins. Eq.~(\ref{ham}) is therefore modified to
\begin{equation}
E = -\sum_{\langle i, j \rangle} J_{ij} S_i S_j  - \sum_i h_i S_i ,
\label{rfham}
\end{equation}
where we have allowed for a different field $h_i$ on each site.
The random field model is obtained if one chooses
the $h_i$
at random with zero average value, and has the $J_{ij}$ unfrustrated (so we
could set them all to equal unity). Again
there are experimental systems which have been widely studied but which space
does not allow me to discuss. For more information see Ref.~\cite{bely} and the
articles in Ref.~\cite{book}.
In Eq.~(\ref{rfham}) frustration comes from competition between
the interactions on the one hand,
which prefer the spins to be parallel, and the random
fields on the other, which prefer the spins to follow the local field
direction.

The traditional physics approach to studying problems with frustration and
disorder is the Monte Carlo simulation, i.e. a random sampling of the states
according to the Boltzmann distribution~\cite{boltz}. However, at low
temperatures the system gets trapped in one of the valleys for a long ``time''
and is only very rarely able to escape over a barrier in the energy landscape
to another valley. The probability for escape is exponentially small in the
ratio of the barrier height to the temperature. As a result, equilibrium
simulations can only be done on very small systems at low temperatures. Some
speed up can be obtained from recently developed Monte Carlo algorithms such
as parallel tempering~\cite{hn,kpy} (also known as ``exchange Monte Carlo'')
but the range of sizes that can be studied is still quite limited. 

In this talk I will discuss an alternative approach which uses sophisticated
optimization algorithms from Computer Science~\cite{heiko}
to find the exact ground
state.  The idea will be to ``beat the small size limit'' of Monte
Carlo methods.  The advantages of the computer science approach are:
\begin{enumerate}
\item
It is exact. There are no statistical errors or problems of equilibration.
\item
One can study large sizes.
\end{enumerate}
However, there are also some disadvantages. These are:
\begin{enumerate}
\item
Only the ground state is determined, so one is restricted to
zero temperature properties.
\item
Only for some models are there efficient algorithms.
\end{enumerate}

In the rest of this talk I will discuss what has been learned from applying
optimization algorithms to the spin glass and random field
problems, and also describe some prospects for the future.

\section{The Random Field Model}
In this section I will discuss how optimization algorithms have
enhanced our understanding of the random field model.
The energy is given by
Eq.~(\ref{rfham}) with the sites on a regular lattice, which we take to be a
square grid in two-dimensions, a simple cubic grid in three-dimensions, and
similarly in higher dimensions. The interactions $J_{ij}$ are all set to unity
and the random fields are chosen from a symmetric distribution with mean and
variance given by
\begin{equation}
[ h_i] = 0, \qquad [h_i^2] = h_R^2,
\end{equation}
where the rectangular brackets
$[\cdots]$ denote an average over the disorder, so
$h_R$ is the strength of the random field.

In the absence of random fields it is known that there is a non-zero
magnetization $\langle S_i \rangle$ at low temperatures and  we say that there
is ``long range order''. This long range order vanishes
continuously at critical temperature. When the random fields are turned on one
could ask whether even a small random field prevents
the formation of long range order at {\em any}\/ temperature or whether a
critical field strength is needed to destroy long range order at low
temperature. A famous argument due to Imry and Ma~\cite{im} states that for
dimension two and lower, the random field always ``wins'' in the sense that
long range order is destroyed by an arbitrarily weak random field, with the
system ``breaking up'' into domains of parallel spins. The domain size
diverges as $h_R \to 0$ so one recovers long range order for $h_R$ strictly
zero. However, in dimension, $d$, greater than two an arbitrarily small random
field does not cause the system to break up into domains on long length scales
and long range order is preserved up to a critical field strength. 

\begin{figure}
\centerline{\psfig{figure=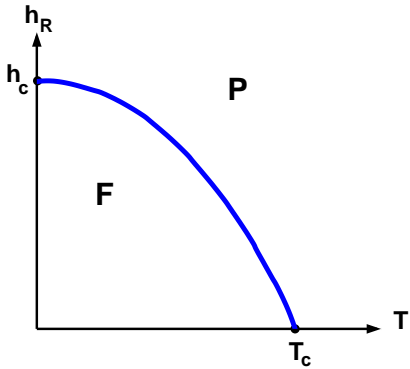,height=5cm}}
\caption{
The phase diagram of the random field model in dimension greater than
two.``P'' denotes the paramagnetic phase with no long range ferromagnetic
order, while ``F'' denotes the ferromagnetic phase. 
}
\label{fig:rfim_pd}
\end{figure}

For $d > 2$, the
phase diagram is that sketched in Fig.~\ref{fig:rfim_pd}. For $h_R
= 0$ the ferromagnetic phase disappears at $T = T_c$ due to thermal
fluctuations, while at $T=0$ the ferromagnetic phase disappears at a critical
value of the random field, $h_c$, due to the disordering
effects of the random field.  This will be important later.

What aspects of the random field problem are physicists interested in? It
turns out that many quantities vary with a power law in the vicinity of the
critical point. Denoting by $\delta$ the deviation from the phase boundary in
Fig.~\ref{fig:rfim_pd}, the magnetization varies, for $\delta$ small, like
\begin{equation}
\langle S_i \rangle \sim |\delta|^\beta ,
\end{equation}
where $\beta$ is known as a ``critical exponent''.  Other quantities of
interest are the specific heat, $C$, and the magnetic susceptibility, $\chi$,
which vary as
\begin{eqnarray}
\langle C \rangle & \sim & \delta^{-\alpha} , \nonumber \\
\langle \chi \rangle & \sim & \delta^{-\gamma} . 
\end{eqnarray}
Critical exponents such as $\alpha, \beta$ and $\gamma$ are of interest
because they are universal~\cite{univ}, only depending on broad features of the
problem such as the dimensionality of the lattice and whether or not there is
a random field. They are not expected to depend on the strength of the random
field, as long as it is non-zero, or the distribution used for the random
fields, as long as it is symmetric.

Physicists would like to the understand universal critical behavior, such as
the values of the critical exponents.
Since the phase boundary is crossed at zero temperature by varying $h_R$ one
{\em can}\/ investigate the behavior near the phase boundary
using optimization (i.e. $T = 0$) algorithms. Fortunately, determining the
ground state energy is equivalent to a Max-Flow problem~\cite{heiko} which can
be solved in polynomial time. This was first realized by Barahona and 
Angl\`es d'Auriac et
al.~\cite{ada} and subsequently used by Ogielski~\cite{og},
Sourlas and collaborators~\cite{sourlas}, and others. 

Here I will just discuss briefly some of the results found in
Ref.~\cite{sourlas}, which investigated the random field model in $d=3$.
In their optimized implementation of the Max-Flow algorithm, they studied $L^3$
lattices up to $L = 90$, and found empirically that the CPU time varied as
$L^4$. This is remarkably efficient, being not much more than the time ($L^3$)
needed to
scan once through the lattice. Ref.~\cite{sourlas} provides strong
evidence that the transition is actually discontinuous, corresponding to an
exponent $\beta = 0$. This had been suspected earlier from finite-$T$
Monte Carlo simulations~\cite{heiko_mc} on sizes up to $L=16$
but the results of Ref.~\cite{sourlas}
are more convincing because they are on much larger systems. Normally a first
order transition leads to a latent heat and rather weak fluctuation effects
compared with a continuous transition. However, no latent heat is seen in
the random field problem and, in other respects, there seem to be large
fluctuation effects characteristic of a continuous transition. This
dichotomy is not understood. 

Several different types of random field distribution were used in
Ref.~\cite{sourlas}. While they all gave rise to $\beta = 0$, other
quantities, also expected to be universal, seemed to depend on the type of
disorder, casting some doubt on the hypothesis that universality holds for
random systems. This important question needs further work.

The $T=0$ approach cannot easily determine the specific heat exponent
$\alpha$, which is unfortunate because there is a discrepancy between the
experimental value~\cite{bkj}
which is close to zero and results from Monte Carlo
simulations, e.g. Ref.~\cite{heiko_mc}, which give $\alpha \simeq -0.5$.

\section{The Spin Glass}
The energy of the spin glass problem is given by Eq.~(\ref{ham}) where the
$J_{ij}$ are taken from a symmetric distribution with mean and variance given
by
\begin{equation}
[ J_{ij}] = 0, \qquad [J_{ij}^2] = 1.
\end{equation}
One often takes a Gaussian distribution, though another
popular choice is
is the bimodal
distribution, also the called $\pm J$ distribution, in which the interactions
have values $+1$ and $-1$ with equal probability. The latter distribution
has the special
feature that there are {\em many}\/ ground states (we say that the ground
state is ``degenerate''). In fact the number of ground states is exponentially
large in the number of spins $N$ giving rise to a finite ground state entropy. 

Two principal questions have been asked about spin glasses:
\begin{enumerate}
\item
Is there a phase transition at finite temperature $T_c$?
\item
What is the nature of the spin glass state below $T_c$?
\end{enumerate}

For the first question, Monte Carlo simulations and early (unsophisticated)
ground state calculations have shown;
\begin{itemize}
\item
In $d=2$ the transition is only at $T=0$.
\item
In $d=3$ (and higher) the transition is at finite temperature.
\end{itemize}
The conclusion for $d=2$ is very strong and so is the situation in $d=4, 5,
\cdots$ etc. The case of $d=3$ has been the most difficult to resolve and
earlier work was not very conclusive, but the most recent
simulations~\cite{sue} seem rather convincing.

Concerning the second question,
we have already noted that, because of the complicated
energy landscape, there are large clusters of (carefully chosen) spins which
can be flipped with rather low energy cost. Is it possible to quantify this
remark? Two principal scenarios have been
proposed which differ, mainly, as to the energy of
these large-scale excitations. These scenarios are:
\begin{itemize}
\item
The ``droplet model'' of Fisher and Huse~\cite{fh}. In this phenomenological
picture a few very plausible assumptions are made. The lowest
energy to create an excitation of linear size $L$ is assumed to vary as
\begin{equation}
\Delta E \sim L^\theta ,
\label{de_droplet}
\end{equation}
where $\theta\ (> 0)$ is an exponent. $\theta$ can not be negative if $T_c >
0$ otherwise
there would be large scale excitations which cost vanishingly small energy and
the system would break up into domains at any finite temperature. We shall see
below that this is what
happens in $d=2$. Note that for a ferromagnet there is a
positive energy cost for each interaction on the wall of the excitation and,
since the wall area goes like $L^{d-1}$, one has $\theta = d-1$ in that case.
However, for a spin glass it turns out that $\theta < d-1$ (in fact it is also
true that $\theta < (d-1)/2$). Hence, there is a near cancellation between
the effects of the bonds which were ``unsatisfied'' before the excitation is
flipped and then become ``satisfied'' (which lower the excitation energy),
and the the effects of the satisfied bonds which become unsatisfied (which
increase the excitation energy).

\item
The ``replica symmetry breaking'' (RSB) picture of Parisi~\cite{parisi}. The
Parisi theory is the (presumably) exact solution of an artificial model with
infinite-range interactions. The assumption is then made that qualitatively
similar behavior also occurs for more realistic models with short range
interactions. An important ingredient of the RSB picture is that there are
excitations of order the size of the system whose energy does {\em not}\/
grow with the size of the system i.e.
\begin{equation}
\Delta E \sim \mathrm{const.}
\end{equation}
This is in contrast to the prediction of the droplet theory in
Eq.~(\ref{de_droplet}). The cancellation between the effects of the satisfied
and unsatisfied bonds on the boundary of the excitation
is then even more complete than in the
droplet model.

\end{itemize}

To discuss what has been learned from optimization algorithms it is necessary
to distinguish $d=2$ from higher dimensions. We first consider $d=2$.

We have already noted that a
square is frustrated if an odd number of its bonds are negative and the
converse, that the square is unfrustrated (i.e. each bond can be satisfied) if
there are an even number of negative bonds, is also true.
Changing the sign of the bonds in such a way that the frustration remains
unchanged has no effect on the ground state energy because it can be
compensated for by changing the sign of appropriate spins. 
Hence, for the $\pm J$ distribution, the
ground state energy is determined entirely by the location of the
frustrated squares. For a distribution in which the magnitude of the bonds is
not constant we also need to keep track of the magnitude of the bonds (though
not the sign) plus the location of the frustrated squares. 

\begin{figure}
\centerline{\psfig{figure=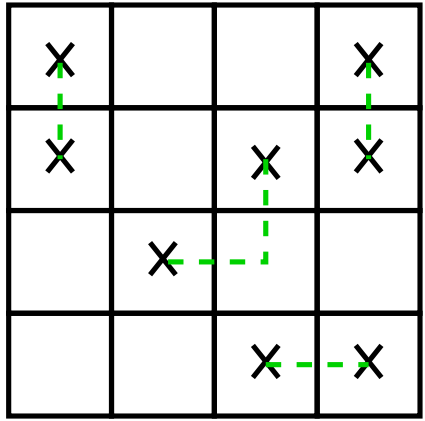,height=6cm}}
\caption{
A two dimensional spin glass in which the frustrated squares are denoted by
crosses. The ground state energy is obtained from the minimum length of the
strings connecting the frustrated squares. For a distribution of interactions
where the
magnitude as well as the sign varies there is a weight to
each string segment equal to the magnitude of the bond which it crosses.
}
\label{fig:frust_sq}
\end{figure}

Let us therefore indicate the frustrated squares on the lattice by drawing a
cross in their center, as shown in Fig.~\ref{fig:frust_sq}. We indicate
the unsatisfied bonds by drawing dashed 
lines at right angles across them. Clearly
the lines must begin and end on frustrated squares and so form ``strings''
connecting the crosses. The ground state energy is increased relative to that
of the unfrustrated system by the energy of the bonds crossed by the strings.
For a $\pm J$ distribution the ground state energy is therefore determined by
minimizing the length of the strings. For a distribution of interactions where
the magnitude as well as the sign varies one has to minimize the total
``weight'' of the string, where the weight of a string segment is equal to the
magnitude of the bond which it crosses.

This problem
is equivalent to a Minimum Weight Perfect Matching Problem~\cite{heiko} as
first realized by Barahona et al.~\cite{barahona}
which can be solved in polynomial
time, i.e. it belongs to the category ``P'' of optimization problems. To be
more precise it is a polynomial algorithm provided the lattice is a ``planar
graph'', i.e. it can be drawn on a piece of paper with no lines crossing.
Unfortunately this rules out periodic boundary conditions which are often
imposed to eliminate surface effects which arise
from the spins on the surface having a
different number of neighbors from the spins in the bulk.
With periodic boundary conditions the
problem belongs to the class ``NP''. However, an efficient ``Branch and Cut''
algorithm~\cite{heiko} enables quite large sizes to be studied~\cite{juenger}.

In three dimensions or higher calculating the ground state of spin glass is NP
for all boundary conditions. Most work has used ``heuristic'' algorithms,
which are not guaranteed to give the exact ground states, but which, when used
carefully, do seem to give the true ground state in most instances. The most
effective such approach seems to be the ``genetic algorithm'' developed for
spin glasses by Pal~\cite{pal} and subsequently used by Palassini and
me~\cite{py1,py2} and Marinari and Parisi~\cite{mp}.

Let us now discuss what has been learned about spin glasses from optimization
techniques, starting with $d=2$.

\begin{figure}
\centerline{\psfig{figure=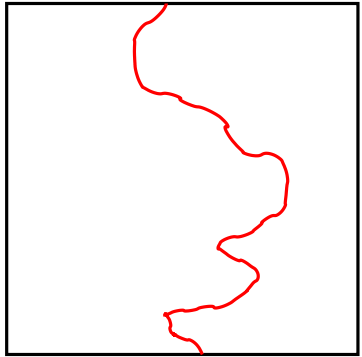,height=5cm}}
\caption{
A domain wall crossing a two-dimensional lattice. Its energy will be of order
$L^\theta$, where is $L$ is the size of the lattice in each dimension, and it
will have a fractal dimension $d_s$.
}
\label{fig:wall}
\end{figure}

First we note that the restriction to $T=0$ is more serious than
for the random field problem because we cannot go through the critical point.
We can, however, learn about low energy excitations by computing the ground
state, then perturbing the system in some way, and finally recomputing the
ground state. As an example let us start with periodic boundary
conditions and then change to anti-periodic boundary conditions in one
direction, which simply corresponds to changing the sign of the bonds across
one boundary. This induces a domain wall across the system, as shown in
Fig.~\ref{fig:wall}, such that all the spins on one side of the wall are
flipped.

The wall will have an energy, which could have either sign, and its
characteristic scale varies with the size of the system in each direction, $L$,
as in Eq.~(\ref{de_droplet}). Starting with the pioneering work of Bray and
Moore~\cite{bm}, and
followed by later studies~\cite{rj,py2d} for larger sizes (up to about
$30^2$) using the Branch and Cut method, it has been found that $\theta$
is negative in $d=2$ with a value of around $-0.28$. The negative value means
that the system will break up into large domains at any finite temperature,
so there is no spin glass state except at $T=0$. These studies also show that
the wall is a fractal with fractal dimension $d_s$ about 1.28, greater than 1
so it is not a smooth curve, but also less than 2 so the wall is not ``space
filling''.

Recently Middleton~\cite{midd} has used the Matching algorithm to determine
ground states of two-dimensional spin glasses with free boundaries for very
large sizes, up to $512^2$, using a different approach to generate excitations
from which $\theta$ and $d_s$ can be determined. His results agree well with
the other work.

The case of $d=3$ is more interesting that that of $d=2$ not only because this
is the physical dimension but also because there is a finite temperature spin
glass state.
There is general agreement that $\theta \simeq 0.20$ in $d=3$, starting with
the first studies of Bray and Moore~\cite{bm} which could only consider sizes
up to $4^3$, and followed by later work~\cite{py1,hartmann} which could go up to
about $10^3$ using heuristic optimization algorithms. The positive value
indicates that the spin glass state should be stable at low but finite
temperatures. Ref.~\cite{py1} also find that $d_s \simeq 2.68$.

Subsequently, two papers~\cite{km,py2} have argued that a different value of
$\theta$, consistent with
zero, is obtained from excitations in which the boundary
conditions are not changed but a carefully chosen set of spins is flipped, for
example by thermal noise. This suggests that the spin glass is actually quite
close to the RSB  picture. However, the sizes are still quite small, up to
around $10^3$, and the assertion that there are two values (at least) for
$\theta$ depending the type of excitation being considered is quite messy, so
this claim needs further study.

Given the considerable interest in the spin glass in three-dimensions, it is
unfortunate that there are no polynomial algorithms for finding the ground
state. It is to be hoped that, in the future, algorithms will be developed
which both give exact ground states and can treat larger sizes than the present
heuristic algorithms.

\section{Conclusions}
This discussion of the role of optimization algorithms in statistical physics
has been very brief. For further information the reader should consult the
references. Ref.~\cite{heiko} is a good place to start. 

The main conclusions of this talk are:
\begin{enumerate}

\item
Algorithms from computer science have ``broken the size barrier'' for some
problems in statistical physics, e.g.
\begin{enumerate}
\item
The random field model
\item
The spin glass in two-dimensions
\end{enumerate}

\item
The application of algorithms from computer science to physics problems works
best as a collaboration between computer scientists and physicist, e.g.
Ref.~\cite{rj}.

\item
For the future I expect there will be developments in the following areas:
\begin{enumerate}
\item
More models will be solved.
\item
More efficient algorithms will be developed for NP problems such as the spin
glass in three-dimensions. So far, with the genetic algorithm, we can study up
to of order $10^3$ spins. Surely we do better than this.
\item
Spin glasses will be used to investigate the statistics of ``hardness''. For
example, given an algorithm for the exact ground state of a three-dimensional
spin glass such as branch-and-cut, one can study the {\em distribution} of
CPU times required to solve the problem for different realizations of
disorder. It would be interesting to see how the {\em average}\/ CPU time
varies with system size and do the same for
the {\em typical}\/ (e.g. median) CPU time.
If the distribution of CPU times is very broad, the average may be
dominated by a few rare samples which are extremely ``hard'' and vary with
size in a different way from the typical CPU time. This distinction has been
made recently in statistical physics in the study of some {\em quantum}\/
systems undergoing phase transitions at zero temperature~\cite{fisher} but, to
my knowledge, does not seem to have been investigated systematically in
studies of hardness of NP problems. 
\end{enumerate}

\end{enumerate}

\medskip
\noindent {\bf Acknowledgments:}\\
I would like to thank Heiko Rieger for educating me about the algorithms
mentioned in this talk and many other topics. Much of my own work in this
field has been with Matteo Palassini and I would like to
thank him for a stimulating collaboration.
I am especially grateful to the organizer, Reinhard Wilhelm, for
inviting me to speak at the 10th Anniversary Dagstuhl conference, which
introduced me to large areas of computer science about which I knew nothing
before. My research is supported by the NSF under grant DMR-9713977.


\begin{thebibliography}{99}
\bibitem{boltz}
    According to statistical mechanics, a system in thermal equilibrium has a
    probability proportional to $\exp(-E_l/k_B T)$ of being in a state $l$
    with energy $E_l$, where $T$ is the temperature, and $k_B$ is Boltzmann's
    constant (usually set to unity in model calculations). This exponential is
    known as a ``Boltzmann factor''.
\end{thebibliography}
\end{document}